\documentclass[preprint,aps,showkeys,showpacs]{revtex4-1}   
\usepackage{epsfig}
\usepackage{xcolor}   
\usepackage{amsmath}

\begin{document}

\title{Modulational instability and quantum droplets
in a two-dimensional Bose-Einstein condensate}

\author{Sherzod R. Otajonov}
\author{Eduard N. Tsoy}
\author{Fatkhulla Kh. Abdullaev}

\address{Physical-Technical Institute of the Uzbek Academy of Sciences,\\
Chingiz Aytmatov Str. 2-B, Tashkent, 100084, Uzbekistan}

\begin{abstract}
Modulational instability of a uniform two-dimensional binary Bose-Einstein condensate (BEC)
in the presence of quantum fluctuations is studied. The analysis is based on the coupled
Gross-Pitaevskii equations. It is shown that quantum fluctuations can induce instability
when the BEC density is below a threshold. The dependence of the growth rate of modulations
on the BEC parameters is found. It is observed that an asymmetry of the interaction
parameters and/or initial densities of the components typically decreases the growth rate.
Further development of the instability results in a break-up of the BEC into a set of quantum
droplets. These droplets merge dynamically with each other so that the total number of
droplets decreases rapidly. The rate of this decrease is evaluated numerically for
different initial parameters.
\end{abstract}

\maketitle

\section{Introduction}
\label{sec:intro}

   Modulation instability (MI), or the Benjamin-Fair instability~\cite{Benjamin1967,Ostrovskii1967},
is a well-known phenomenon in physics. The main effect of MI is an exponential growth of
the modulation amplitude of a plane wave. MI has been studied in different fields of
physics, such as nonlinear optics, plasma physics, hydrodynamics, and the physics of
Bose-Einstein condensates (BECs), see e.g.~\cite{Hasegawa1975,Agarwal2019,Salasnich2003}. A
usual method for studying MI in BECs is based on the linear stability analysis of a
stationary plane-wave. The dynamics of MI in the presence of additional effects has been
considered in many works. These works include a study of MI of matter-waves under a strong
periodic variation in time of the scattering length~\cite{Abdullaev2009}, MI in chiral BECs
with zero-energy nonlinearity~\cite{Bhat2021}, and the influence of three-body interaction
on MI in coupled nonlinear systems~\cite{Baizakov2018}. A nonlinear stage of the MI has
also been studied ~\cite{Zakharov2013,Vanderhaegen2021}.

   According to the mean-field theory, MI in single-component BECs occurs only for the
attractive two-body atomic interaction. This is because the attractive interaction
counteracts a condensate expansion caused by the quantum pressure. The presence of MI in
two-component BECs depends on the relationship between the values of the parameters of
interactions~\cite{Goldstein1997, Kourakis2005}. For example, the attractive inter-component
interaction might induce MI, even when both components have the self-repulsive interaction.
When both intra- and inter-species interactions are repulsive, MI can still occur due to
the development of out-of-phase structures in the components~\cite{Kourakis2005}.

   A study of MI is important because the instability is a precursor to the formation of spatially
localized patterns, such as bright solitons and self-bound quantum droplets (QDs).
Recent theoretical studies of 1D binary BEC have been reported in Refs.~\cite{Mithun2020}
where MI results in the formation of multiple QDs.

    A quantum droplet is a structure with a localized distribution of the
condensate density. In contrast to bright solitons that can have an arbitrary peak density,
the peak density of QDs is limited.
The peak density of a condensate in QDs is small for a small number of atoms, and it saturates at a
large number of atoms ~\cite{Petrov2015, Petrov2016, Astrakharchik2018, Otajonov2019,
Li2018, Otajonov2020}. Therefore, the properties of QDs with a large number of atoms are
similar to those of incompressible liquids. A possibility of a formation of QDs in BECs due to
quantum fluctuations were predicted theoretically in Refs.~\cite{Petrov2015,
Petrov2016}.

   The effect of quantum fluctuations is small compared with the two-body interaction.
Both the quantum fluctuation parameter and the two-body interaction parameter
depend on the scattering length, and therefore they cannot be varied
independently via the Feshbach resonance.
Quantum fluctuations can be made noticeable by using, for example, two-component or
dipolar condensates. In these systems, one can tune the parameters of atomic interactions
independently of the strength of quantum fluctuations. In a binary BEC, parameters of the
interaction between atoms can be chosen such that there is a small residual attraction.
This attraction can result in a collapse of a BEC. However, quantum fluctuations that
introduce a repulsion can prevent collapse. In dipolar BECs, the strength of the dipolar
interaction and the scattering length can be tuned independently, such that the effect of
quantum fluctuations becomes perceptible.

The static and dynamical properties of QDs in 1D without external confinement are
investigated theoretically in Refs.~\cite{Astrakharchik2018,Otajonov2019}. Exact solutions,
collisions of QDs, and the period of oscillations
of a breathing mode are found. Two different regimes of QDs depending
on the QD size or the number of atoms are also identified.
In the 2D case, stability regions of QDs with embedded vorticity are found numerically in
Ref.~\cite{Li2018}. For analytical treatment of 2D QDs and vortices, the variational
approximation is proposed in Ref.~\cite{Otajonov2020}.
In 3D, properties of two-component QDs and vortices are reported in Ref.~\cite{Kartashov2018},
stability regions are found for vortices $S$=1 and 2, where $S$ is the topological charge.
The dynamics of QDs under the periodic variation of scattering parameters is studied in Ref.~\cite{Otajonov2022}.
Collective oscillations of the Lee-Huang-Yang fluid are investigated in Ref.~\cite{Jorgensen2018}.

The existence of QDs has been confirmed experimentally in different physical systems, such as
single-component BECs with dipole-dipole interaction~\cite{Barbut2016,Barbut2018},
homonuculear~\cite{Cabrera2018,Skov2021}, and heteronuclear~\cite{D'Errico2019} bosonic mixtures.
These theoretical and experimental studies have opened new perspectives
for the investigation of fundamental properties of quantum gases under the
action of quantum fluctuations~\cite{Barbut2019,Luo2021}.

   We consider in this work a system with a contact interaction and quantum fluctuations. The present
study aims to investigate the linear and nonlinear stages of MI, and the formation of QDs in a binary 2D BEC.
The structure of the paper is as follows. In Sec.~\ref{sec:model} we introduce the model equations for
the description of a binary 2D BEC. The linear stability analysis is presented in Sec.~\ref{sec:linearMI}.
Stability/instability regions are found for different values of the parameters.
The nonlinear stage of MI is analyzed in Sec.~\ref{sec:nonlinearMI}.
The number of emerging QDs is estimated, and its variation in time is analyzed using image processing
methods. In Sec.~\ref{sec:conc} we summarize our findings.

\section{The model and results}
\subsection{The model}
\label{sec:model}

   In Ref.~\citep{Petrov2016}, the energy density of two-dimensional binary BECs
in the presence of quantum fluctuations is derived. Also, the Gross-Pitaevskii equation is
obtained via a standard variation of the energy density
functional. In 2D, the contribution of quantum fluctuations depends
logarithmically on the condensate density, and this is the specific property of atomic
scattering in two dimensions. The dynamics of a two-component BEC under the action of
quantum fluctuations is described by the coupled Gross-Pitaevskii
equations~\cite{Petrov2016}, see also ~\cite{Li2018, Otajonov2020}:
\begin{eqnarray}
& i \hbar \, \cfrac{\partial \Psi_j}{\partial T}+ \cfrac{\hbar^2 }{2 m} \nabla ^{2} \Psi_{j}
-\cfrac{ g_{j} m } { 4 \pi \hbar^2 } P_j \log \left[ \cfrac{e m}{ \hbar^2 \Delta} P \right] \Psi_{j} -
\nonumber \\
& \sqrt{g_{j}} \left( \sqrt{g_{j}} |\Psi_{j} |^{2} - \sqrt{g_{3-j}} |\Psi_{3-j} |^{2} \right ) \Psi_{j}  =0,
\label{dgpe}
\end{eqnarray}
where $\Psi_j$ is a wave function of the component, $j=1$ and $2$, $\nabla^2=\partial^2_X+\partial^2_Y$
is the two-dimensional Laplacian, $T$ is the time, $m$ is the atomic mass,
$P=g_{1} |\Psi_{1}|^{2} + g_{2} |\Psi_{2}|^{2}$, $g_j=4 \pi \hbar^2  \sigma_j / m$ is the intra-species interaction
parameter, $\sigma_j= 1/\log[4 e^{-2 \gamma} /a_j^2 \Delta]$,
$
\Delta = \cfrac{4 e^{-2 \gamma} }{ a_{12} \sqrt{a_1 a_2}}$ $ \exp
   \left[ \cfrac{- \log^2(a_2/a_1) }{ 2 \log[a_{12}^2 / (a_1 a_2)]} \right]
$, parameters $a_1, a_2$ and $a_{12}$ are the 2D scattering lengths, and $e$ and $\gamma$ are the Euler
number and the Euler constant, respectively. Following Ref. ~\cite{Petrov2016}, we consider
the case of the weak intra-species repulsion ($g_1,g_2> 0$) and inter-species attraction
($g_{12}<0$). The parameter $g_{12}$ is taken such that $g_{12}^2=g_1 g_2$, this relation
is reflected in the definition of parameter $\Delta$. Equation~(\ref{dgpe}) shows that when
the interaction parameters ($g_1, g_2, g_{12}$) tend to zero, the effect of quantum fluctuations ($\sim
g^2$) is negligible. By a proper choice of the interaction parameters and the component
densities, it is possible to make the strength of quantum fluctuations [the third
term in Eq.~(\ref{dgpe})] comparable with the two-body interaction [the last term in Eq.~(\ref{dgpe})].
Also, notice that quantum fluctuations in 2D result in a self-attraction for small $P$ (the
re-scaled density) and a self-repulsion for large $P$~\cite{Petrov2016}.

By using new variables, $\psi=\Psi/\psi_s$, $t=T/t_s$ and $(x,y)=(X,Y)/r_s$, where scale
parameters $\psi_s$, $t_s$ and $r_s$ are defined as:
\begin{eqnarray}
 & \psi_s=\left( \Delta / 8 \pi e \sqrt{\sigma_1 \sigma_2} \right)^{1/2} \,, \qquad
 t_s=m e / \hbar \Delta \sqrt{\sigma_1 \sigma_2} \, ,
\nonumber \\
 &r_s= ( e/  \Delta \sqrt{\sigma_1 \sigma_2} )^{1/2},
\label{rescale}
\end{eqnarray}
Eq.~(\ref{dgpe}) is reduced to the following dimensionless form:
\begin{eqnarray}
 & i\, \cfrac{\partial \psi_j}{\partial t}+ \cfrac{1 }{ 2} \nabla ^{2} \psi_{j} - \cfrac{\psi_{j}}{2 \sqrt{\sigma_1 \sigma_{2} }}
    \left( \sqrt{ \sigma_j / \sigma_{3-j} }\, |\psi_{j} |^{2} - |\psi_{3-j} |^{2} \right )  -
\nonumber \\
 &  \sqrt{ \sigma_j / \sigma_{3-j} }\, \psi_{j}\, p \log (p) = 0,
\label{gpe}
\end{eqnarray}
where $\nabla^2=\partial^2_x+\partial^2_y$ and $p=(\sigma_{1} |\psi_{1}|^{2} + \sigma_{2} |\psi_{2}|^{2}) / (2 \sqrt{\sigma_1 \sigma_2})$.
All theoretical and numerical results are for the dimensionless equation. However, in Sec.~\ref{sec:nonlinearMI} we also provide parameters in physical units.
A uniform distribution of the condensate is described by a plane wave solution of Eq.~(\ref{gpe}):
\begin{equation}
  \psi_{j}=A_{j}\exp(-i \, \mu_{j} \, t ) \, ,
\label{cw}
\end{equation}
where $A_j$ and $\mu_j$ are the amplitude and chemical potential of the $j$-th component, respectively.
The dependence of the chemical potentials on the amplitudes is found from Eq.~(\ref{gpe}) and (\ref{cw}):
\begin{eqnarray}
  &\mu_{j}=\cfrac{A_j^2}{2 \sigma_{3-j}}-\cfrac{A_{3-j}^2}{2 \sqrt{\sigma_j \sigma_{3-j}}} +
  \sqrt{\sigma_j / \sigma_{3-j}}\, p \log (p).
\label{omegaj}
\end{eqnarray}
In the following Sections, we study the growth of modulations of the uniform state (Sec.~\ref{sec:model} B),
and the formation of QDs in the later stage of instability (Sec.\ref{sec:model} C).

\subsection{The linear stage of MI}
\label{sec:linearMI}

  For the linear stability analysis, we study the dynamics of small perturbations
$\delta\psi_{j}\ll A_{j}$ imposed on the stationary state
\begin{equation}
  \psi_{j}=(A_{j}+\delta\psi_{j})\exp(-i \, \mu_{j} \, t) \, .
\label{pert}
\end{equation}
The small-amplitude dynamics of $\delta \psi_j$ is described by the following equations:
\begin{eqnarray}
  & i \, \cfrac{\partial \delta \psi_j}{\partial t}+{1 \over 2} (\delta\psi_{jxx}+\delta\psi_{jyy})-c_{j}(\delta\psi_{j}^{*} +
\delta\psi_{j}) -
\nonumber \\
& c_{3}(\delta\psi_{3-j}^{*}+\delta\psi_{3-j})=0,
\label{linear}
\end{eqnarray}
where $j=1$ and $2$,
\begin{equation}
  c_{j} \equiv \cfrac{A_j^2}{2 \sigma_{3-j}} + \cfrac{A_j^2 \sigma_j}{2 \sigma_{3-j}} \log \left( e p_0 \right)  \, ,
\label{omeg1}
\end{equation}
\begin{equation}
  c_3 \equiv \cfrac{A_1 A_2}{2 } \left[- \cfrac{1}{\sqrt{\sigma_1 \sigma_2} } + \log \left( e p_0 \right) \right]  \, ,
\label{omeg2}
\end{equation}
and $p_0=(\sigma_{1} A_1^2 + \sigma_{2} A_2^2) / (2 \sqrt{\sigma_1 \sigma_2})$.

We represent the perturbation as $\delta \psi_j= u_j+i\, v_j$, and split Eq.~(\ref{linear})
into the real and imaginary parts. Assuming that $(u_j,v_j) \sim \exp{(\lambda t+ik_x x +i
k_y y})$ we get the following characteristic equation:
\begin{eqnarray}
   \lambda^4 + \lambda^2 k^2 \left( c_1+c_2+\frac{k^2}{2} \right) + \frac{k^8}{16} +
\nonumber \\
      \frac{k^6}{4}\left( c_1 + c_2 \right) + k^4 (c_1 c_2 -c_3^2)=0 \, ,
\label{lam4}
\end{eqnarray}
where $k^2=k_x^2+k_y^2$.

Equation~(\ref{lam4}) is a bi-quadratic equation on $\lambda$, and its solution is
\begin{equation}
  \lambda_{\pm}^{2}={k^2 \over 4} \left[ -k^2 -2 (c_1+c_2) \pm 2 \sqrt{(c_1-c_2)^2+4c_3^2} \right] .
\label{sol2}
\end{equation}

   A plane wave is modulationally unstable if the following conditions are fulfilled:
\begin{equation}
   e \, p_0 < 1,  \ \ \mathrm{and\ \ } |k| < k_{cr},
\label{logarg}
\end{equation}
where the critical value $k_{cr}$ is defined as:
\begin{equation}
  |k| < k_{cr} \equiv \sqrt{2} \sqrt{-(c_1+c_2)+\sqrt{(c_1-c_2)^2+4c_3^2}}.
\label{kcr}
\end{equation}
For $\sigma_1=\sigma_2$, parameter $p_0$ is equal $n_0/2$. Therefore, condition
(\ref{logarg}) states that MI occur at sufficiently low densities. The real part of
exponents, $G \equiv \mathrm{Re}(\lambda_\pm) > 0$, characterizes the growth rate of MI. If
$e \, p_0 > 1$, the system is modulationally stable. When $e \, p_0=1$,  terms with
logarithm, associated with a contribution from quantum fluctuations, in Eqs.~(\ref{omeg1})
and ~(\ref{omeg2}) become zero. We recall that the presence of MI in a binary system
without quantum fluctuations depends on relative values of intra- and inter-species
interaction parameters~\cite{Kourakis2005}. For our choice of $g_1$, $g_2$ and $g_{12}$,
the system without quantum fluctuations is neutrally stable. This means that the growth rate
is zero for these parameters, and the system is modulationally stable. Quantum fluctuations
induce a self-attraction for low densities. This can result in the emergence of MI and QDs
in the system. We mention that QDs in binary BECs have been observed experimentally in
Refs.~\cite{Cabrera2018, Skov2021, D'Errico2019}, where the role of quantum fluctuations was
revealed.

   The maximum value of the MI growth rate $G_{max}$ is attained at the corresponding $k = k_{max}$, where
\begin{equation}
G_{max}={k_{max}^2 \over 2}={k_{cr}^2 \over 4} \, , \qquad  k_{max}={k_{cr} \over \sqrt{2}}\,.
\label{maxkG}
\end{equation}
In the symmetric case, $\sigma_1 = \sigma_2 = \sigma$ and $A_1 = A_2 = A$, equations for MI
parameters $G$ and $k_{cr}$ are simplified, because $c_1 = c_2 = A^2 [1/\sigma +
\log(eA^2)]/2$  and $c_3 = A^2 [-1/\sigma + \log(eA^2)] / 2$. We analyze the dynamics for
different sets of parameters. In all cases studied, it is found that  for a given total
density $n_0 = n_{10} + n_{20} \equiv A_1^2 + A_2^2$, parameter $G_{max}$ is larger for the
symmetric case, compared with asymmetric cases. In particular, when the density of one component
is small, the corresponding $c_j$ and $c_3$ are also small. Then, $k_{cr}$ is negligible,
assuming $c_1, c_2 >0$, so MI exists in a narrow region of $k$, with the low growth rate,
see Eq.~(\ref{maxkG}).

\begin{figure}[htbp]
  \centerline{ \includegraphics[width=4.5cm]{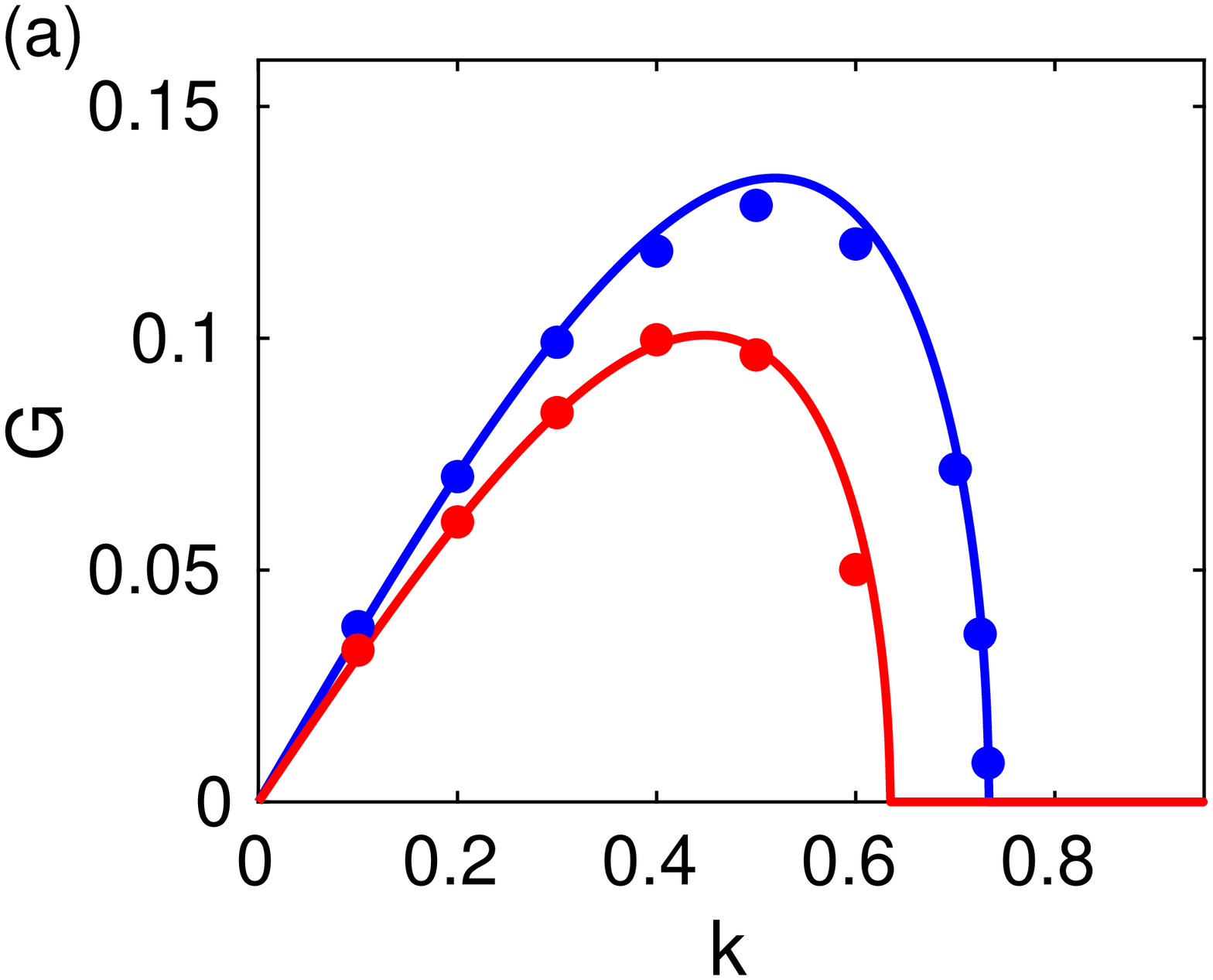} \hskip-0.4cm
     \includegraphics[width=4.5cm]{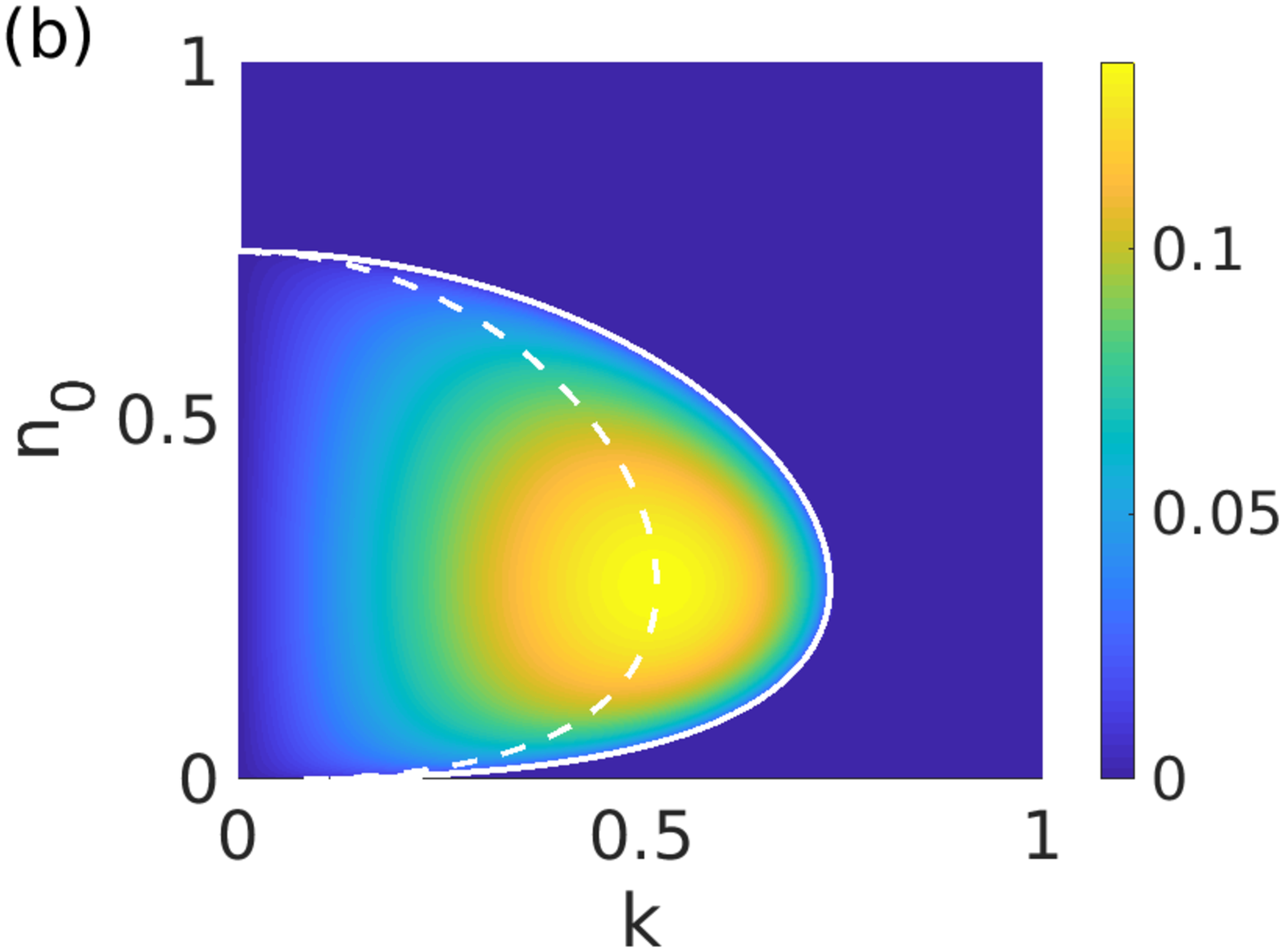}}
\caption{(a): The MI growth rate $G$ vs $k$. The top (blue) line for the symmetric case $n_{10} = n_{20} =0.3$
and the bottom (red) line for the asymmetric case $n_{10} = 0.3$ and $n_{20} = 0.1$. Points are
found from direct numerical simulations of Eq.~(\ref{gpe}). (b): The growth rate $G$ as a
function of $k$ and $n$ for the symmetric case. The solid line
shows the MI boundary, see Eq.~(\ref{kcr}), and the dashed line represents $G_{max}$, see
Eq.~(\ref{maxkG}). The interaction parameters are $\sigma_1=\sigma_2 = 0.1$.}
\label{fig-1ab}
\end{figure}

   In Fig.~\ref{fig-1ab}(a), the MI growth rate as a function of the modulation wave number $k$ is plotted for
the symmetric (a top line) and the asymmetric (a bottom line) cases. The growth rate profile has a
typical (half-) butterfly shape. Points in Fig.~\ref{fig-1ab}(a) show the results of
numerical simulations. The dependence of $G$ on $k$ and $n_0$ for the symmetric case is
presented in Fig.~\ref{fig-1ab}(b). One can see that MI exists only if the BEC total
density is sufficiently small. This is due to the repulsive nature of quantum fluctuations
at large densities. We mention that MI and the formation of chains of QDs in 1D binary BEC
were analyzed theoretically in Ref.~\cite{Mithun2020}.

\begin{figure}[htbp]
  \centerline{ \includegraphics[width=4.3cm]{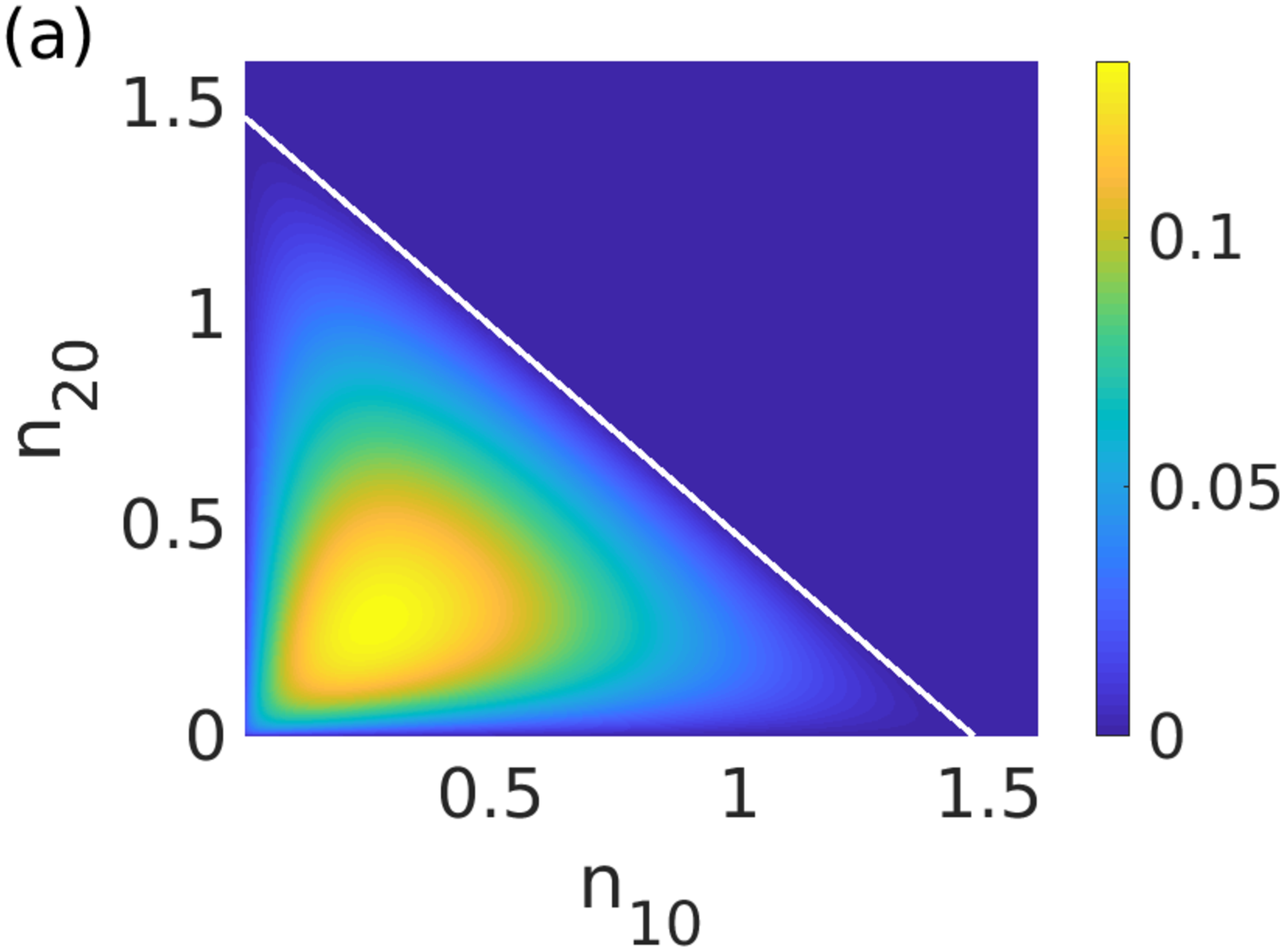} \hskip-0.08cm
   \includegraphics[width=4.3cm]{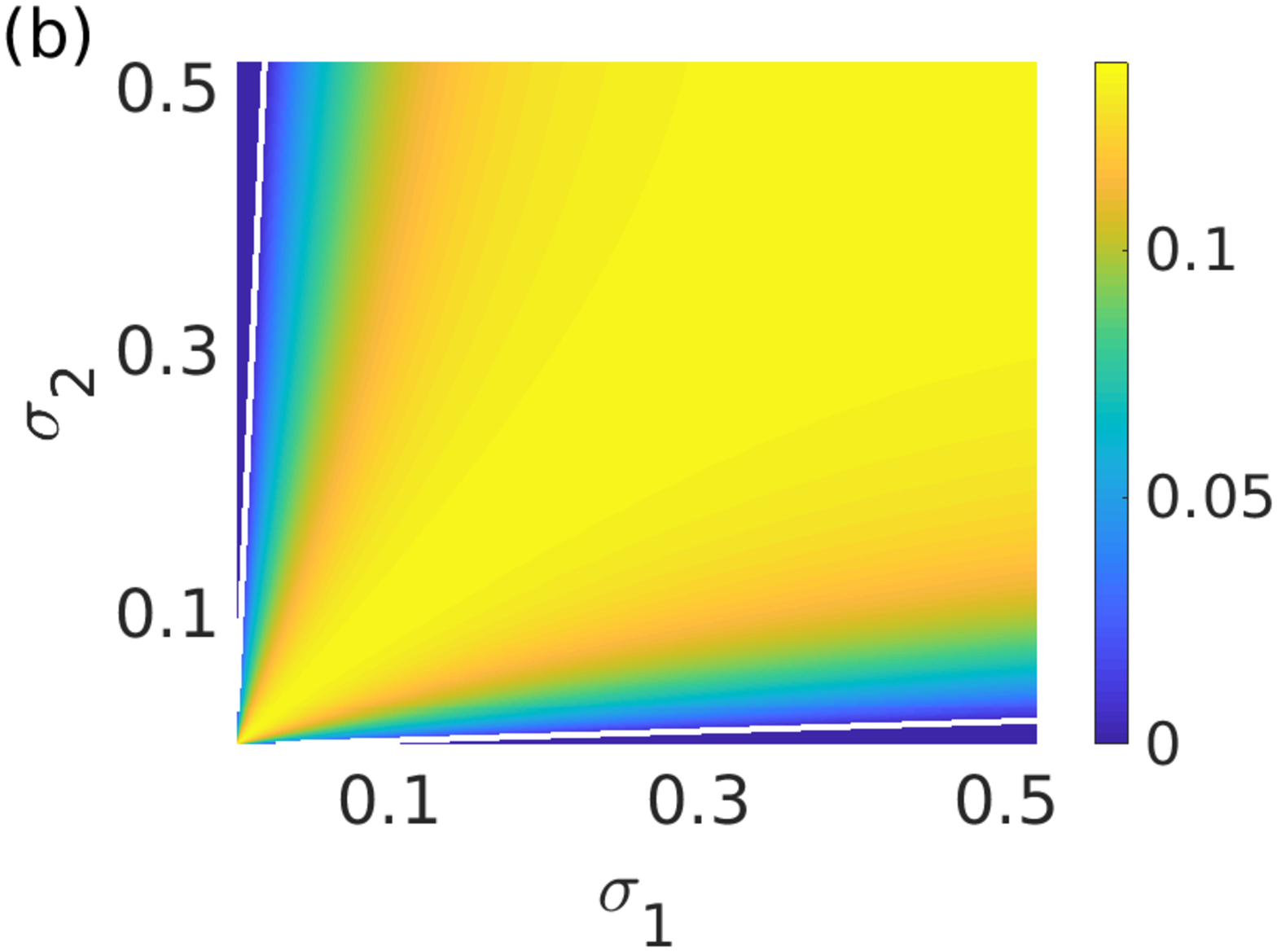} }
\caption{(a) The growth rate of MI in $(n_{10}, n_{20})$-plane for $\sigma_j=0.1$.
Maximum of the growth rate equals $0.135$ at $n_{10} = n_{20} = 0.271$.
(b) The growth rate in $(\sigma_1, \sigma_2)$-plane for $n_{10} = n_{20} = 0.271$. In both figures, lines represent
the instability boundaries, see Eq.~(\ref{kcr}).
}
\label{fig-2ab}
\end{figure}

   An influence of asymmetry of the BEC parameters on the MI growth rate is shown in Fig.~\ref{fig-2ab}.
The rate $G$ as a function of initial amplitudes $(n_{10}, n_{20})$ for given $\sigma_1=
\sigma_2 = 0.1$ is shown in Fig.~\ref{fig-2ab}(a). The instability region is below $n_{20}=
2/e - n_{10}$ line, see Eq.~(\ref{logarg}). Maximum of $G$, which equals to 0.135 for such
values of $\sigma_1$ and $\sigma_2$, is realized also for the symmetric case.

   In Fig.~\ref{fig-2ab}(b), the instability region in ($\sigma_1, \sigma_2$)-plane is between
two straight lines $\sigma_2=\sigma_1 A_2^{-4} (- A_1^2 A_2^2 + 2 /e^2 \pm (2 / e)\sqrt{1 /e^2
-A_1^2 A_2^2})$. Figure~\ref{fig-2ab}(b) is plotted for $n_{10} = n_{20} = 0.271$, a value
that corresponds to the maximum gain 0.135 in Fig.~\ref{fig-2ab}(a). For a given $n_{10} =
n_{20}$, the maximum value of the growth rate attains at the diagonal $\sigma_2=\sigma_1$,
and it does not depend on the values of $\sigma_j$ on that line. We conclude from
Fig.~\ref{fig-2ab} that asymmetry of the system parameters and/or of the component
densities results typically in a decrease of the MI growth rate.

   Our analysis suggests the following way to observe  MI in experiments. Firstly, one needs
to create a high-density two-component condensate. For this condition, the condensate is
stable. Then, a decrease of the density $n$, for example, by expanding an external trap,
can induce MI when condition~(\ref{logarg}) is satisfied.

   Due to the instability, the uniform state is transformed into a structure of peaks
and dips. These peaks can be associated with strongly overlapped QDs. The distance between
the peaks is $\sim 2 \pi /k_{max}$. Therefore, the number of QDs per unit area is evaluated
as $\rho_0 = K_0/L^2$, where $K_0 = k_{max}^2 L^2 / (2 \pi)^2$ is the number of generated
QDs. We mention that parameters $K_0$ and $\rho_0$ are valid for the linear stage of MI,
when $t<t_{th}$. The threshold time $t_{th}$ can be estimated as the time when the
modulation amplitude is of order $\sim 0.1 A_0$, $t_{th}  \simeq G_{max}^{-1} \log(0.1
A_0/\epsilon)$, where $\epsilon$  is the initial amplitude of modulations.

   In order to check the dynamics, we perform numerical simulations of Eq.~(\ref{gpe})
in the domain $-50 \leq L(=L_x=L_y) \leq 50$, with $1024 \times 1024$ grid points. The size
of the domain is much larger than the typical size of droplets, which emerged in the linear
stage of MI ($L \gg 1/k_{max}$). We use the split-step Fourier transform method with
periodic boundary conditions. We employ two types of initial conditions. The first type of
initial conditions is a plane wave with periodic modulations $\delta \psi (x,y,0) =
\epsilon \, \cos(k_x x) \cos(k_y y)$. The second type is a plane wave with random
modulations. In experiments, the instability is typically induced from random
perturbations, consisting of modes in a wide range of $k$. All unstable $k$ contribute to
the growth of modulations. However, the wave mode with $k = k_{max}$ that corresponds to
the maximum of gain dominates in the dynamics. We observe that by using random modulations,
the dynamics do not depend strongly on initial conditions in different runs, see
Sec.\ref{sec:nonlinearMI}.

In order to obtain the growth rate for a particular $k$, we use periodic modulations with
$\epsilon=10^{-3}$. We monitor the dependence of the maximum modulation amplitudes on time
and recover the value of the growth rate $G$. Points, found from numerical simulations of
Eq.~(\ref{gpe}), in Fig.~\ref{fig-1ab}(a) match well with the prediction of the linear
theory. We also mention that for regular perturbations with given $k$, the density
distribution almost returns to its initial stage with very small modulations. This
resembles a well-known MI recurrence phenomenon~\cite{Infeld1981}. However, after 2-5
cycles of returning to the initial stage, an interaction between droplets occurs.

\subsection{The nonlinear stage of MI}
\label{sec:nonlinearMI}

   We use numerical simulations in order to analyse the development of MI. As an initial
condition, we use a noisy plane wave $\psi_j(x,y,0)=A_j[1+ \epsilon R(x,y)]$, where
$\epsilon \ll 1$, and $R(x,y)$ is a random function with the uniform distribution of values
in a range $(- 1, 1)$.

\begin{figure}[htbp]
   \centerline{ \includegraphics[width=4.4cm]{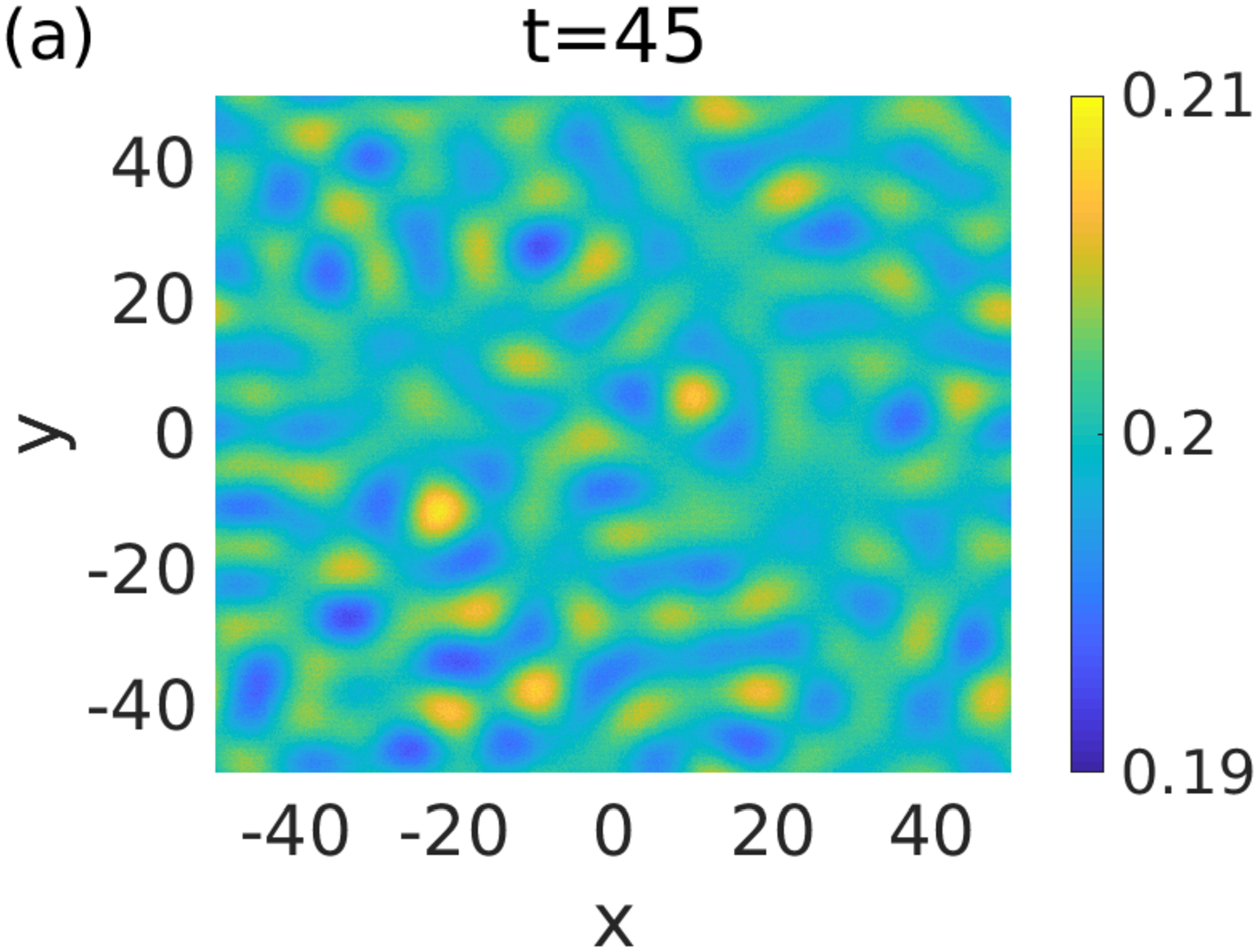} \hskip-0.1cm
      \includegraphics[width=4.4cm]{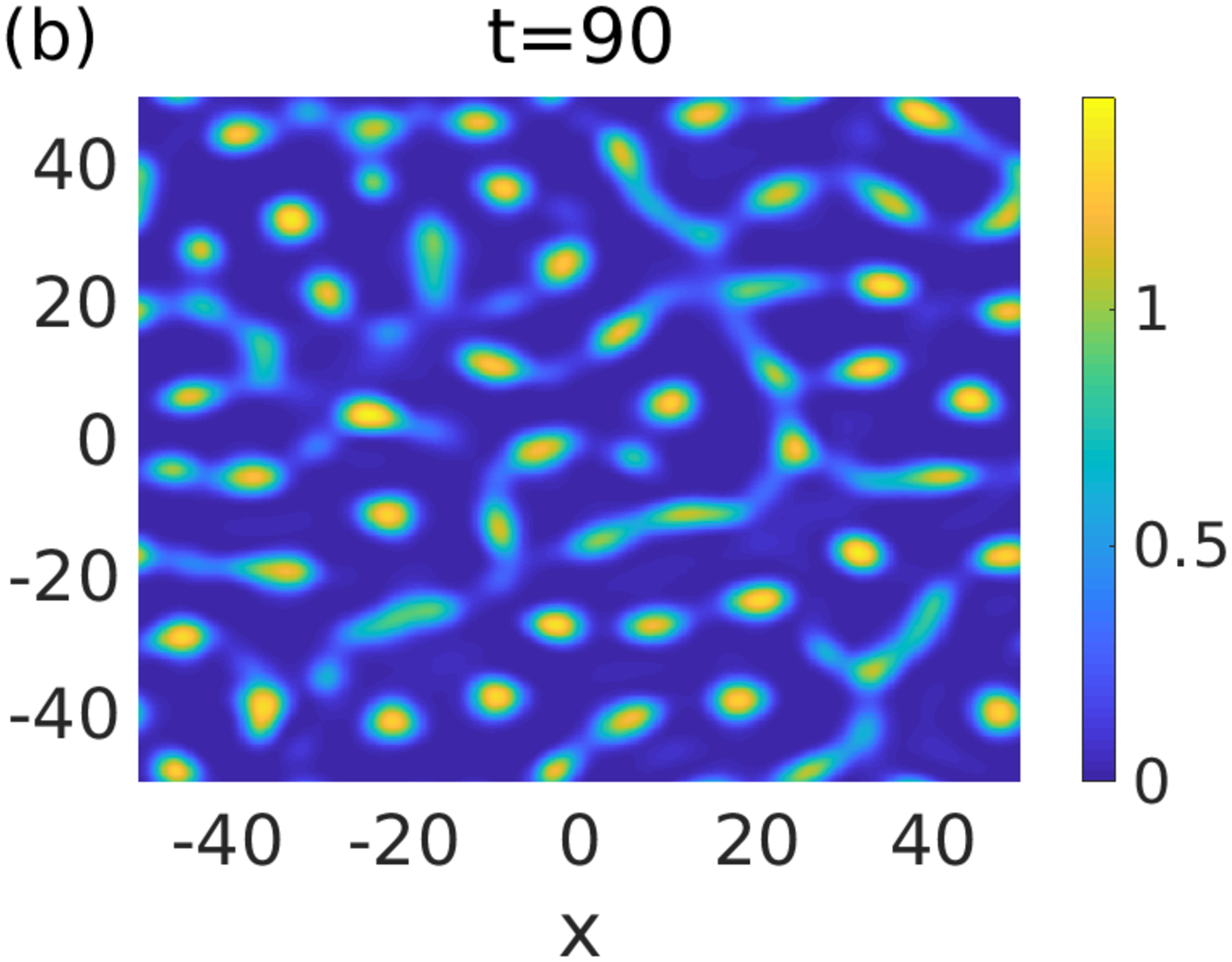}}
   \centerline{ \includegraphics[width=4.4cm]{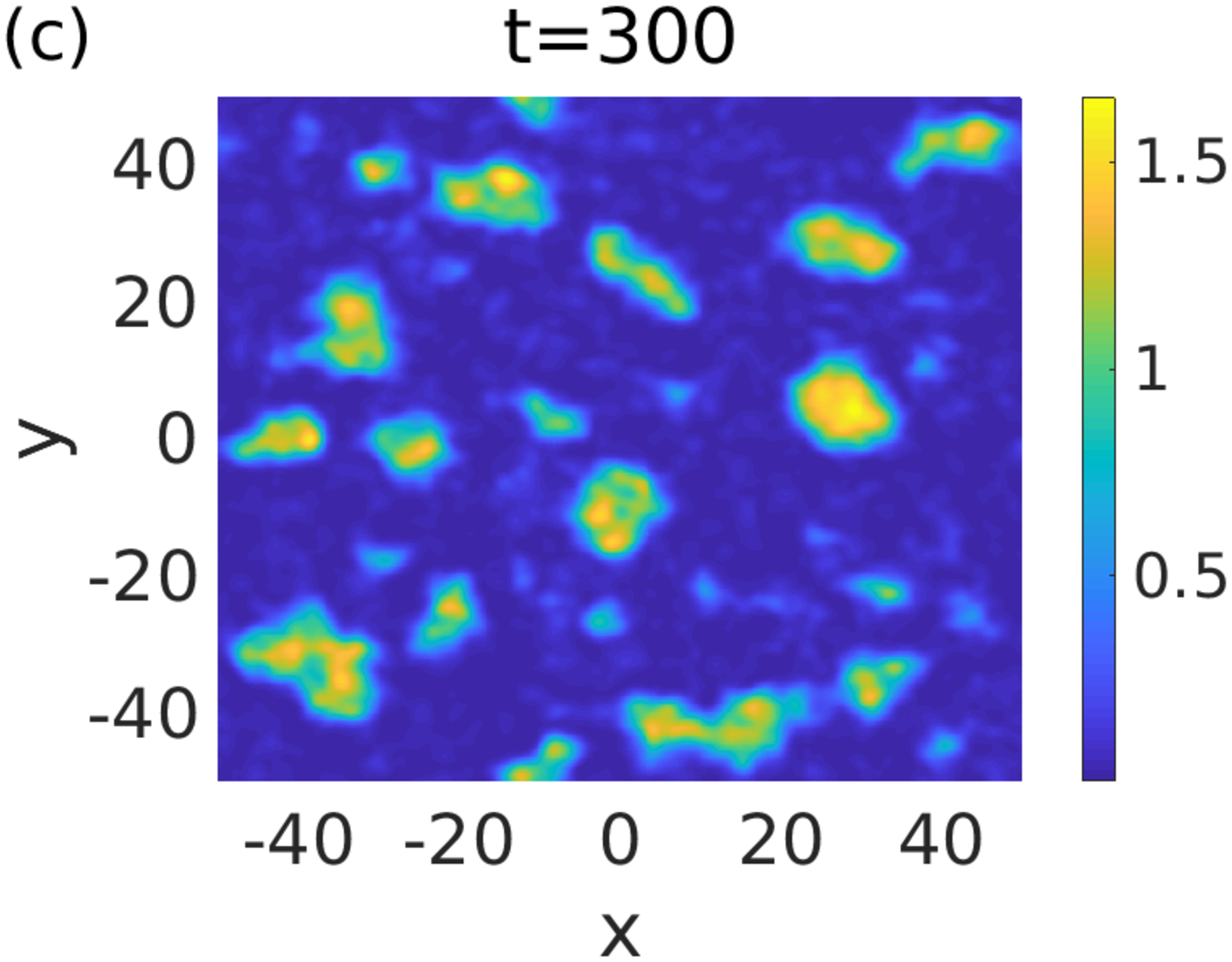} \hskip-0.1cm
      \includegraphics[width=4.4cm]{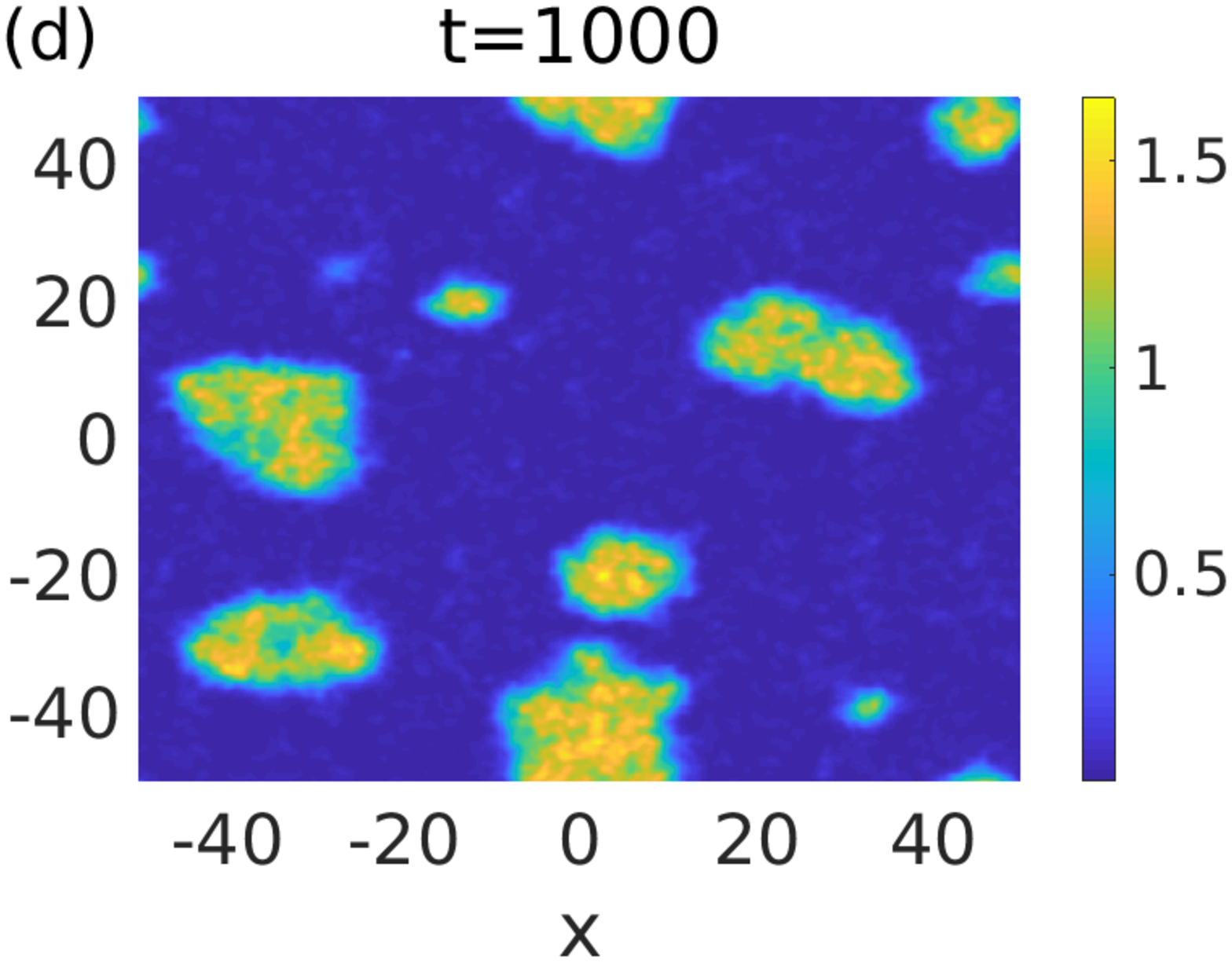}}
\caption{The typical dynamics of the density distribution $n=|\psi_1^2|+|\psi_2|^2$ at (a) $t=45$, (b) $t=90$,
(c) $t=300$, and (d) $t=1000$. The initial densities are $n_0=0.2$, and $\sigma_1=\sigma_2=0.1$.}
\label{fig-5abcd}
\end{figure}

\begin{figure}[htbp]
  \centerline{ \includegraphics[width=6.cm]{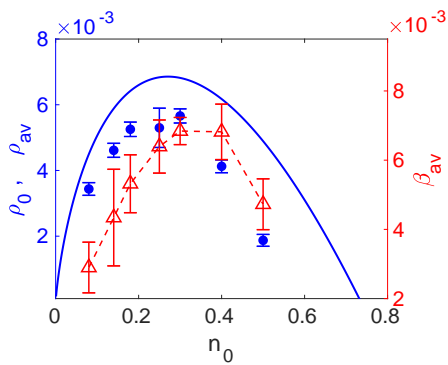} }
\caption{ The number of generated QDs per unit area $\rho_0$ (solid line), found from the MI theory,
the averaged number $\rho_{av}$ (circles), found from numerical simulations of Eq.~(\ref{gpe}),
and the averaged decrement $\beta_{av}$ of the QD number (triangles, right axes) vs
the initial density $n_0$.}
\label{fig-3}
\end{figure}

   Figures~\ref{fig-5abcd}(a) and (b) show typical patterns of the overlapped droplets
developed in the linear stage of MI. In Fig.~\ref{fig-5abcd}(a), the density distribution
still corresponds to a modulated plane wave with $n(x, y, t)$ near the initial value $n_0$
(notice a different scale on this subplot).
This means that there is no fragmentation of the BEC into droplets in Fig.~\ref{fig-5abcd}(a).
Figure~\ref{fig-5abcd}(b) shows a result of a plane wave break-up into distinguishable QDs
such that intervals between them have zero density. The mean numbers of generated QDs, found
over several simulations, on this stage are in a qualitative agreement with the value $K_0$, obtained
from the linear analysis, see also Fig.~\ref{fig-3}. One can see that some QDs are located
close to each other, forming continuous clusters. The distance between QDs is close to the
size of droplets. Therefore there is a strong interaction between QDs that causes their
merging.

   A process of the merging results in the creation of larger QDs, as seen in
Fig.~\ref{fig-5abcd}(c) and (d). Since the shape of QDs is far from a stationary form, and
due to the merging, there are strong oscillations of particle amplitudes and widths.
Moreover, the density distribution within a single droplet is also non-uniform and varies on
time. However, we do not observe a noticeable movement of QDs within the plane. We mention
that a similar merging (coalescence) of droplets in a BEC with the cubic-quintic interaction
was reported in Ref.~\cite{Josserand1997}.

\begin{figure}[htbp]
\centerline{ \includegraphics[width=4.5cm]{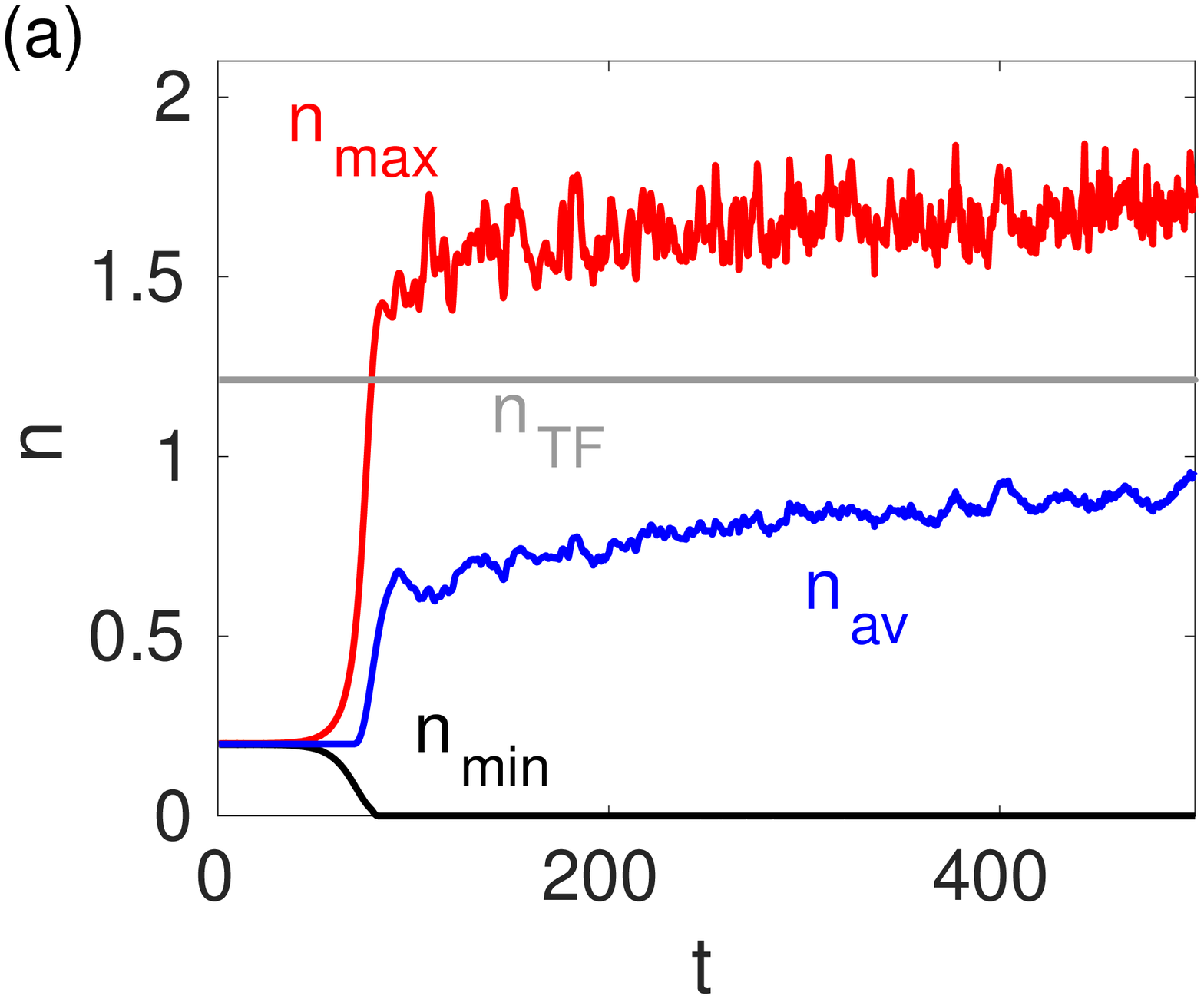} \hskip-0.39cm
\includegraphics[width=4.5cm]{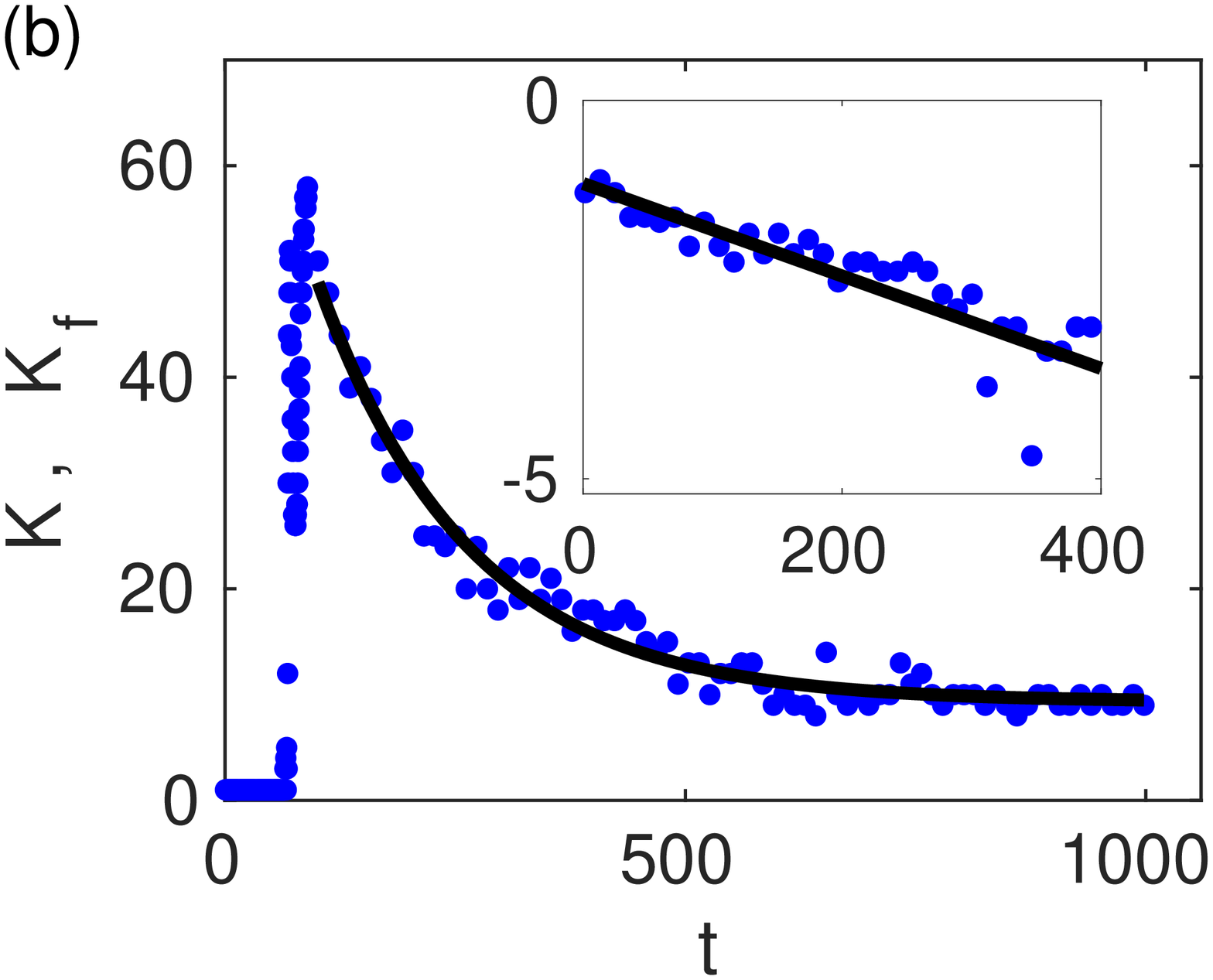} }
\caption{(a) Time evolution of $n_{max}$ (top line), and $n_{min}$ (bottom line),
$n_{av}$ (middle line) for $n_0 = 0.2$, and $\sigma_1 = \sigma_2 = 0.1$.
The straight line represents the Thomas-Fermi limit $n_{\mathrm{TF}}$
(b) The number $K$ of generated QDs on time. Points are found from direct numerical simulations,
the line is found from the exponential regression.
The parameters of fitting are $a=75.28$, $\beta=0.0057$, $t_b\simeq 90$, $b=8.57$. The inner plot shows this relation in logarithmic scale. The horizontal and vertical axis are $t-t_b$ and $\log[(K-b)/a]$, respectively.
}
\label{fig-4ab}
\end{figure}

  We measure numerically the maximum $n_{max} \equiv $ $\max_{x,y} [n(x,y,t)]$, the minimum $n_{min} \equiv \min_{x,y} [n(x,y,t)]$, and the average $n_{av}$ of the BEC density, where $n(x, y,t) \equiv |\psi_1|^2 + |\psi_2|^2$, see Fig.~\ref{fig-4ab}(a). Parameter $n_{av}$ is found within QDs, i.e. we exclude points where $n(x,y,t) < 0.5\, n_{max}$.
One can see that at $t \gtrsim 90$, the system is in the nonlinear stage. After a rapid increase, $n_{max}$ varies near a constant value. This value corresponds to spikes of BEC density oscillations of different QDs. Average density $n_{av}$ of QDs is well below $n_{max}$. This density can be estimated from the following reasoning. As it is found in
Ref.~\cite{Li2018, Otajonov2020}, the peak density of quantum droplets saturates at a large number $N_{QD}$ of particles within a single droplet. This fact reflects the liquid nature of QDs. As demonstrated in Ref.~\cite{Otajonov2020}, the peak density of a stationary
QD for the symmetric case can be found approximately from
\begin{equation}
  n_{st} = 2 \exp \left(-\frac{1}{2} + \frac{1}{2m} - \frac{2^{1/m} \pi m }{N_{QD}} \right),
\label{nst}
\end{equation}
where the form parameter $m$ is defined as $m = (0.4433 + 0.05906 N_{QD})^{0.5047}$ for $N_{QD} = [1,1000]$.
Equation~(\ref{nst}) is derived, assuming a super-Gaussian profile of a QD, i.e. $\psi(x,y,t) \sim \sqrt{n_{st}}\, \exp[-(r/w)^{2m} / 2]$.
Factor 2 in front of the exponent in Eq.~(\ref{nst}), compared with the corresponding equation in Ref.~\cite{Otajonov2020}, accounts for two components.
Stationary amplitude  $n_{st}$ tends to the Thomas-Fermi limit~\cite{Li2018} $n_{TF} = 2 / \sqrt{e}$, when $N_{QD} \to \infty$. As it follows from Fig.~\ref{fig-4ab}(a), average density $n_{av}$ indeed approaches on time to the Thomas-Fermi limit. During the development of MI, maximum droplet density $n_{max}$ rises, while $n_{min}$ vanishes. This corresponds to a break up of a plane wave into separated QDs (c.f. Fig.~\ref{fig-5abcd}(a) and (b)).

  As follows from Eq.~(\ref{logarg}), in the symmetric case, plane waves with
$n_0 > 2/e$ are stable. At the same time, the peak density $n_{st}$ of QDs is changed
from 0 at $N_{QD}= 0$ to $n_{TF} = 2 / \sqrt{e}$ at $N_{QD} \to \infty$. This suggests
that QDs with large $N_{QD}$ should be stable against small modulations, while those
with small $N_{QD}$ can be unstable. However, in Ref.~\cite{Li2018,Otajonov2020} it was
shown that QDs with an arbitrary number of particles are stable. We suppose that small-amplitude
QDs are stable due to the quantum pressure.

  When the initial densities are not equal, the scenario of the droplet formation is different.
Let us consider, for example, the case $A_1 > A_2$. The density maxima $n_{max,1}$ and
$n_{max,2}$ of the components are increased on time during the interaction of QDs, while
the minima $n_{min,1}$ and  $n_{min,2}$ are decreased, similarly to the case of equal
initial amplitudes, c. f. Fig.~\ref{fig-4ab}. This process of the sharp variation ends when
 the density minimum $n_{min,2}$ reaches zero. After that, parameters $n_{max,1}$, $n_{max,2}$
and $n_{min,1}$  are changed gradually, with random fluctuations. In regions between
droplets, the density of the first component $n_1(x, y) \sim n_{min,1}$, while $n_2(x, y)
\sim 0$. However, emerged droplets consist of the condensate mixture. Therefore, at later
stages, the density distribution corresponds to a set of two-component droplets immersed in
a single-component background. This is similar to droplets of water in a cloud of vapor.

  We recall the instability is suppressed in the absence of one component. Further
development of MI on the  background is ceased or is decelerated due to a negligible value
of the density of the second component in regions between droplets. We mention that this
type of dynamics is observed when the difference between the initial densities is
appreciable (more than 10-20\%). If the initial densities of the components are close to
each other, the dynamics is similar to the symmetric case.

  The number $K(t)$ of QDs is found from the density distribution analysis.
Namely, we transform the density distribution $n(x, y,t)$ to a black-and-white image in
$(x,y)$-plane, using $0.5\, n_{max}(t)$ as a threshold, and we count the number of
connected components (islands). We mention that a choice of a different threshold (e.g.
$0.7\,n_{max}$) does not change strongly the results obtained. For small times, $K(t)$
equals one, since at each point, $n(x, y,t) > 0.5\,n_{max}(t)$. When instability is
developed, $K(t)$ grows rapidly due to emergence of droplets from small modulations, see
Fig.~\ref{fig-4ab}(b). We mention that this rapid increase can be non-monotonic, that is
value $K(t)$ can oscillates and have local maxima during this growth. Starting from $t =
t_b$, the number of QDs decreases exponentially. Therefore, time $t_b$ corresponds to the
boundary between the rapid growth and the decrease of the number of QDs. This decrease is
due to merging of overlapped droplets. Points in Fig.~\ref{fig-4ab}(b) show a variation
of $K(t)$ in a single run of numerical simulations, and the line represents the averaged
fitting curve.

   We fit the dependence $K(t)$, using the following function
\begin{equation}
   K_{f}(t)= a \exp[-\beta \, (t-t_b) ]+ b, \quad t > t_b.
\label{exp}
\end{equation}
Parameter $a$, $b$ and $\beta$, which is the decrement of the number of QDs, are independent fitting coefficients.
For each realization of an initial condition, we obtain the fitting coefficients.
These coefficients are averaged over ten runs. This averaged parameters are used
for the fitting curve $K_{f}(t)$ in Fig.~\ref{fig-4ab}(b). The inner plot in Fig.~\ref{fig-4ab}(b),
shown on a semi-logarithmic scale, justifies our choice of the fitting function.

   We study numerically  the dynamics of QDs also for different values of $n_0$.
In Fig.~\ref{fig-3}, a solid line shows the number $\rho_0$, found from the MI analysis, of
QDs per unit area versus the initial density $n_0$. Points represent averaged values
$\rho_{av}$ of QDs per unit area, found from numerical simulations of Eq.~(\ref{gpe}) at $t
= t_b$. Time $t_b$ of the beginning of the exponential decrease of $K(t)$ is different in
different realizations of initial conditions. In order to get $t_b$, we find average values
of the number of droplets at each moment of time, obtaining $K_{av}(t)$. We choose the
position of the global maximum of this dependence as time $t_b$, and we use this fixed
value for obtaining $(a, b, \beta)$ in different runs for a given $n_0$. Then, in each run
$K(t_b) = a + b$, so that $\rho_b = (a + b)/L^2$. The average of $\rho_b$ over ten runs is
presented in Fig.~\ref{fig-3}. The vertical error bars correspond to the standard
deviation. A reasonable agreement of the theory of the linear regime and simulations
justifies our approach of counting the QD number. Triangular points, connected with a
dashed line, show the dependence of the average decrement $\beta_{av}$ on  $n_0$.

   For estimation of experimentally relevant parameters, we consider a BEC of $^{39}\mathrm{K}$
atoms in different spin states. In dimensional reduction from 3D to 2D, there is a relation
between two- and three-dimensional scattering lengths, $\vec{a}^{(2\mathrm{D})}= (4 \pi
/B)^{1/2} l_0 \, \exp(-\gamma - \sqrt{\pi/2}\, l_0 / \vec{a}^{(3 \mathrm{D})} )$, where
$B=0.915$, $l_0=(\hbar / m \omega_\perp)^{1/2}$ is the harmonic oscillator length,
$\omega_\perp$ is the radial confinement frequency, see Ref.~\cite{Petrov2016}. For both
dimensions the vector $\vec{a}$ has the following components ($a_{1}, a_{2}, a_{12}$).

   The 2D case is realized when the transverse confinement is sufficiently strong. For
$^{39}\mathrm{K}$, such a confinement is achieved for frequency $\omega_\perp / 2 \pi \sim
10 - 50$ kHz that corresponds to the harmonic oscillator length $l_0 \sim (100-50)$ nm. We
mention that the use of traps with such frequencies is reported in a number of works, see
e.g. Ref. ~\cite{Kramer2005, Gorlitz2001}. Three-dimensional intra- and inter-species
scattering length are taken as $a_{1}^{(3 \mathrm{D})}$ and $a_{2}^{(3 \mathrm{D})} \sim
500\, a_0$, and $a_{12}^{(3 \mathrm{D})}=-(0.9 - 0.99)\, a_{1}^{(3 \mathrm{D})}$, where
$a_0$ is the Bohr radius. The characteristic scales of the system are $r_s \sim (3 - 0.6) \
\mu\mathrm{m}$, $t_s \sim (6 - 0.2)$ ms, $N_s \sim 10 - 2$.  Values of parameter $\sigma_j$
are found to be in the range $\sim (0.06 - 0.15)$. These parameters are achievable in
experiments on BECs.

\section{Conclusions}
\label{sec:conc}

   We have studied modulational instability in two-dimensional binary
Bose-Einstein condensates under the action of quantum fluctuations. Modulational
instability occurs only for low densities when $e \, p_0<1$, otherwise, the system is
stable. Analytical equations for the spectrum of the MI growth rate have been obtained.
Instability regions of perturbed plane waves in the parameter space have been found. The
main peculiarity of MI in a system with quantum fluctuations is the existence of the
threshold density above which the system is stable. It has been demonstrated by means of
numerical simulations that the development of MI results in the formation of droplets.
Interestingly, in the case of unequal initial densities of components, these droplets
are separated by a non-zero background. From the analysis of the nonlinear stage of the MI,
we have obtained the number $K(t)$ of generated QDs. It has been shown that the number of
QDs decreases rapidly due to merging, and the decrease rate $\beta$ of this number has been
estimated. Theoretical predictions are corroborated by the numerical simulations of
governing equations. Our work shows a practical application of an image processing
technique for analyzing the BEC density distribution.

\section*{Acknowledgements}

This work has been funded from the State budget of the Republic of Uzbekistan.


\begin{thebibliography}{99}

\bibitem{Benjamin1967} T. B. Benjamin and J. E. Feir, The disintegration of wave trains on deep water.
1. Theory, J. Fluid Mech. {\bf 27}, 417 (1967).

\bibitem{Ostrovskii1967} L. A. Ostrovskii, Propagation of wave packets and space-time self-focusing in
a nonlinear medium, Zh. Eksp. Teor. Fiz. {\bf 51} 1189 (1966). Transl. Sov. Phys. JETP. {\bf 24},
797 (1967).

\bibitem{Agarwal2019} G. P. Agrawal, Nonlinear fiber optics (Academic Press, 2019).

\bibitem{Hasegawa1975} A. Hasegawa, Plasma instability and nonlinear effects (Springer-Verlag, Heidelberg, 1975).

\bibitem{Salasnich2003} L. Salasnich, A. Parola, L. Reatto, Modulational instability and complex dynamics
of confined matter-wave solitons, Phys. Rev. Lett. {\bf 91}, 080405 (2003).

\bibitem{Abdullaev2009} F. Kh. Abdullaev, A. A. Abdumalikov, R. M. Galimzyanov,
Modulational instability of matter waves under strong nonlinearity management,
Physica D {\bf 238}, 1345 (2009).

\bibitem{Bhat2021} I. A. Bhat, S. Sivaprakasam, and B. A. Malomed,
Modulational instability and soliton generation in chiral
Bose-Einstein condensates with zero-energy nonlinearity,
Phys. Rev. E {\bf 103}, 032206 (2021).

\bibitem{Baizakov2018} B. B. Baizakov, A. Bouketir, S. M. Al-Marzoug, and H. Bahlouli,
Effect of quintic nonlinearity on modulation instability in coupled nonlinear Schr\"odinger systems,
Nonlinear Sciences {\bf 180}, 792, (2018).

\bibitem{Zakharov2013} V. E. Zakharov and A. A. Gelash,
Nonlinear stage of modulation instability, Phys. Rev. Lett. {\bf 111}, 054101 (2013).

\bibitem{Vanderhaegen2021} G. Vanderhaegen, C. Naveau, P. Szriftgiser, A. Kudlinski, M.
Conforti, A. Mussot, M. Onorato, S. Trillo, A. Chabchoub, and N. Akhmediev,
``Extra-ordinary'' modulation instability in optics and hydrodynamics,
Proc. Natl. Acad. Sci. {\bf 118}, e2019348118 (2021).

\bibitem{Goldstein1997} E. V. Goldstein, and P. Meystre,
Quasiparticle instabilities in multicomponent atomic condensates,
Phys. Rev. A {\bf 55}, 2935 (1997).

\bibitem{Kourakis2005} I. Kourakis, P. K. Shukla, M. Marklund, and L. Stenflo,
Modulational instability criteria for two-component Bose-Einstein condensates,
Eur. Phys. J. B \textbf{46}, 381 (2005).

\bibitem{Mithun2020} T. Mithun, A. Maluckov, K. Kasamatsu, B. A. Malomed, and A. Khare,
Modulational instability, intercomponent asymmetry, and formation of quantum
droplets in one-dimensional binary Bose gases, Symmetry \textbf{12}, 174 (2020).

\bibitem{Petrov2015} D. S. Petrov, Quantum mechanical stabilization of a
collapsing Bose-Bose mixture, Phys. Rev. Lett. {\bf 115}, 155302 (2015).

\bibitem{Petrov2016} D. S. Petrov and G. E. Astrakharchik, Ultradilute low-dimensional liquids,
Phys. Rev. Lett. \textbf{117}, 100401 (2016).

\bibitem{Astrakharchik2018} G. E. Astrakharchik, B. A. Malomed, Dynamics of
one-dimensional quantum droplets, Phys. Rev. A {\bf 98}, 013631 (2018).

\bibitem{Otajonov2019} Sh. R. Otajonov, E. N. Tsoy, and F. Kh. Abdullaev,
Stationary and dynamical properties of one-dimensional quantum droplets,
Phys. Lett. A, {\bf 383}, 125980 (2019).

\bibitem{Li2018} Y. Li, Z. Chen, Z. Luo, C. Huang, H. Tan, W. Pang, and B. A. Malomed,
Two-dimensional vortex quantum droplets, Phys. Rev. A \textbf{98}, 063602 (2018).

\bibitem{Otajonov2020} Sh. R. Otajonov, E. N. Tsoy, and F. Kh. Abdullaev,
Variational approximation for two-dimensional quantum droplets, Phys. Rev. E \textbf{102}, 062217 (2020).

\bibitem{Kartashov2018} Y. V. Kartashov, B. A. Malomed, L. Tarruell, and L. Torner,
Three-dimensional droplets of swirling superfluids, Phys. Rev. A {\bf 98}, 013612 (2018).

\bibitem{Otajonov2022} Sh. R. Otajonov,
Quantum droplets in three-dimensional Bose–Einstein condensates,
J. Phys. B: At. Mol. Opt. Phys. {\bf 55}, 085001 (2022).

\bibitem{Jorgensen2018} N. B. J{\o}rgensen, G. M. Bruun, and J. J. Arlt,
Dilute fluid governed by quantum fluctuations, Phys. Rev. Lett. {\bf 121}, 173403 (2018).

\bibitem{Barbut2016} I. Ferrier-Barbut, H. Kadau, M. Schmitt, M. Wenzel, and T.
Pfau, Observation of quantum droplets in a strongly dipolar Bose Gas, Phys.
Rev. Lett. {\bf 116}, 215301 (2016).

\bibitem{Barbut2018} I. Ferrier-Barbut, M. Wenzel, M. Schmitt, F. B\"{o}ttcher, and T. Pfau,
Onset of a modulational instability in trapped dipolar Bose-Einstein
condensates, Phys. Rev. A {\bf 97}, 011604 (2018).

\bibitem{Cabrera2018} C. R. Cabrera, L. Tanzi, J. Sanz, B. Naylor, P. Thomas, P.
Cheiney, and L. Tarruell, Quantum liquid droplets in a mixture of Bose-Einstein
condensates, Science {\bf 359}, 301 (2018).

\bibitem{Skov2021} T. G. Skov, M. G. Skou, N. B. J{\o}rgensen, and J. J. Arlt,
Observation of a Lee-Huang-Yang Fluid, Phys. Rev. Lett. {\bf 126}, 230404 (2021).

\bibitem{D'Errico2019} C. D'Errico, A. Burchianti, M. Prevedelli, L. Salasnich,
F. Ancilotto, M. Modugno, F. Minardi, and C. Fort,
Observation of quantum droplets in a heteronuclear bosonic mixture,
Phys. Rev. Research {\bf 1}, 033155 (2019).

\bibitem{Barbut2019} I. Ferrier-Barbut, Ultradilute quantum droplets, Phys. Today {\bf 72}, 46 (2019).

\bibitem{Luo2021} Z. H. Luo, W. Pang, B. Liu, Y. Y. Li, and B. A. Malomed, A new form of liquid matter:
Quantum droplets, Frontiers of Physics \textbf{16}, 1 (2021).

\bibitem{Infeld1981} E. Infeld, Quantitive theory of the Fermi-Pasta-Ulam recurrence in the nonlinear Schr\"odinger equation, Phys. Rev. Lett. \textbf{47}, 717 (1981).

\bibitem{Kramer2005} M. Kramer, C. Tozzo, and F. Dalfovo, Parametric excitation of a Bose-Einstein condensate in a one-dimensional optical lattice, Phys. Rev. A \textbf{71}, 061602 (2005).

\bibitem{Gorlitz2001} A. G{\"o}rlitz, et al., Realization of Bose-Einstein condensates in lower dimensions,
Phys. Rev. Lett. \textbf{87}, 130402 (2001).

\bibitem{Josserand1997} C.  Josserand and S. Rica, Coalescence and droplets in the subcritical nonlinear Schr\"odinger equation,
Phys. Rev. Lett. {\bf 78}, 1215 (1997).

\end{thebibliography}
\end{document}